# Evolution and Function of SMC Proteins

J. C. Phillips

Dept. of Physics and Astronomy, Rutgers University, Piscataway, N. J., 08854

Abstract

Structural Maintenance of Chromosomes (SMCs) proteins have long rod-like structures immersed in water. Here we use our hydroanalytic methods based on amino acid sequences to discuss their dynamics at multiple length scales identified by evolution. The length scales are ~ 10 – 100 times longer than used in normal studies of sequence evolution. Their hydropathic profiles exhibit many features unique to their structure and function.

## Introduction

Structural Maintenance of Chromosomes (SMCs) are part of a large family of ring complexes that participate in a number of DNA transactions. Their structure is rod-like, with predominantly antiparallel coiled-coil rods connected through a flexible hinge [1]. Here we extend earlier studies of static structures to include dynamical effects associated with hydrostructural domains. Our methods have previously described the evolution of contagiousness of Coronavirus (CoV) quantitatively and in great detail [2-5]. Spikes are the active part of CoV; they are long and slender and immersed in water. SMC rods are similarly immersed in water, and one expects hydroanalysis to yield details of the coiled-coil dynamics mediated by hinges. Conversely, the different ~ 1200 amino acids sequences of SMCs provide large-scale tests of the concepts and methods of hydroanalysis.

Structural maintenance of chromosomes (SMC) proteins are chromosomal ATPases, highly conserved from bacteria to humans, that play fundamental roles in many aspects of higher-order chromosome organization and dynamics. In eukaryotes, SMC1 and SMC3 act as the core of the cohesin complexes that mediate sister chromatid cohesion, whereas SMC2 and SMC4 function as the core of the condensin complexes that are essential for chromosome assembly and segregation. Another complex containing SMC5 and SMC6 is implicated in DNA repair and checkpoint responses [6].

Methods

Our earlier CoV analysis relied on Ψ(aa,W) hydropathic profiles, where Ψ(aa) measures the fractal hydropathicity of each amino acid [6,7]. Protein shapes are economically described by structural domains, which are known in detail for static structures. In general these shapes change slightly when attachment occurs. Here a surprising simplification occurs, which is discussed at length in earlier papers [2,5,8,9]. It is a common experience that the surf of water waves near an ocean shore is larger when there is a strong wind. Similarly, small changes in protein shapes are often driven by waves in water films. It has long been thought in $20^{th}$ century molecular dynamics simulations (MDS) that the interactions of water molecules with individual amino acids are complex, as indeed they are when approached from a multi-parameter Newtonian interatomic force-field perspective. An enormous simplification occurs when the $17^{th}$ century Newtonian view is replace by 18th century thermodynamics and $19^{th}$ century wave descriptions in the $21^{st}$ century structure-based methods described here.

The first step in any hydroanalysis is to establish one or two hydropathicity scales. A given protein can function mainly through either first- or second-order phase transitions, and second order is more common [2-7]. Protein unfolding from water to air is a first-order transition, well-described by the most popular hydropathic scale (KD) [10]. Second-order thermodynamic transitions involve much smaller structural (long-range or allosteric) transitions. Given the lengths of the SMC domains (hundreds of amino acids in coiled coils and hinge), one expects the latter MZ scale [11] to be more effective, while domain interfaces and coiled coil kinks could involve the KD scale.

The next step is to determine a good value for the sliding window width W. Data processing using sliding window algorithms is a general smoothing and sorting technique discussed online. Although evolutionary differences are small in SMC, these small differences can be maximized as an independent way to determine W. By contrast, BLAST and its phylogenetic derivatives always use W = 1, and are unable to connect sequences to dynamical domain motion.

Results: Cohesion

Among SMCs, SMC1A is unique. It encodes a subunit of the cohesin-core complex that tethers sister chromatids together to ensure correct chromosome segregation in both mitosis and meiosis [1]. The topological shape of a globular protein depends strongly on its ins and outs relative to its center. The limits of these ins and outs are associated with hydrophobic and hydrophilic extrema. The range of hydro(ins/outs) differences is defined by the variance [(the mean of squares) - the square of means]/(the square of means), here calculated for a range of W values. After calculating Ψ(aa,W) variances for Human and Chicken SMC1A with the MZ scale, we found their ratio, which measures their shape differences as a function of W. Although these differences are small, they have a clear-cut maximum at W = 57, with the ratio 1.0663. Repeating the same calculation with the KD scale, the maximum falls at W = 63, with the ratio 1.0201 This shows that for SMC1A, the MZ scale is at least three times as effective as the KD scale. BLAST uses W = 1, where the variance ratio is 1.0088, about seven times closer to unity (no evolution) than the MZ value with W = 57. There is also a nearly equal variance maximum ratio at W = 19, which is a subharmonic of W = 57.

The Ψ(aa,W) profiles for Human and Chicken SMC1A with the MZ scale and W = 57 are shown in Fig.1. At first one notices only the close similarity, but looking still more closely, ones sees that the evolutionary differences are concentrated near extrema, and make human extrema either slightly larger (smaller) in the C (N) terminal regions, while leaving the hinge almost unchanged.

The excellent review of SMC1A proteins [1] discusses in its Section 3 the canonical role of cohesion, illustrated in its Fig. 2, which shows the antiparallel coiled-coil structure of the SMC1A-SMC3 heterodimer. In our Fig. 2 the hydroprofiles of these SMC proteins are displayed and discussed in the caption.

In Fig. 3 the hydroprofile of SMC1B is compared to SMC1A. SMC1B is thought to contribute to sister chromatid cohesion, together with SMC3 in [12]. Many experiments have found a variety of structural models [13 ]. While SMC1A and SMC3 show little similarity with BLAST (identity and even positive scores below 40%, with gaps of 15-20%), SMC1B is quite similar to SMC1A (74% positive, with no gaps). Looking at Fig. 3, we notice a large systematic difference: SMC1A is about 15% more hydrophilic. Its unique contributions to cohesion are connected to a stronger interaction with water.

A quantitative measure of the similarities that are obvious in Figs.1-3 is the correlations of these profiles. These are: SMC1A,SMC3: 0.863 (Fig.2), and SMC1A,SMC1B (Fig. 3): 0.930. The latter correlation is much higher than the BLAST positive of 0.74 in part because a constant difference between two profiles alone does not reduce the correlation.

Results: Condensin

SMC2 and SMC4 are part of the condensing complex that organizes and compacts chromosomes. The Human and Chicken ratios of the variances of Ψ(aa,W) as a function of W peak at W = 15 for both SMC2 and SMC4. Compared to SMC1A, the maximum at W = 57 has disappeared, while the maximum W = 19 has shifted to the smaller value of W = 15. The large scale of chromosome segregation was reflected in the large value of W = 57 for cohesion SMC1 and SMC3, whereas the smaller scale of compaction uses the smaller value of W = 15.

The Human profiles for Ψ(aa,15) for both SMC2 and SMC4 are shown in Fig.4. There are ~ 100 aa offsets at the N and C terminals. BLAST finds little similarity. The overlapping regions in Fig. 4 do show a correlation of 0.454. The complex structures of the hinge regions are enlarged in Fig. 5. There are still two peaks, but the spacing in SMC4 has dropped ~ 30 aa, while in SMC2 it is still ~ 80 aa. The existence of two hydrophobic peaks in the hinge regions enables the SMC rods to be aligned more flexibly. The reduction of the separation in SMC4 suggests that in compactions this flexibility is reduced.

Results: DNA Repair

The SMC5/6 proteins play essential roles in the maintenance of genome stability, yet their mode of action is not fully understood [14]. We expect that SMC5 and SMC6 to be more complex than SMC3 and SMC4, just as the latter are more complex than SMC1 and SMC2. As before, we examined Human/Chick variance ratios for both SMC5 and SMC6. The extremum for SMC5 occurs at W = 11, continuing the downward trend with increasing compaction found for cohesions and condensins. There is no extremum for SMC6, so we used W = 11 to plot both Human/Chick profiles.

The Human/Chick SMC5 profiles are shown in Fig. 6, aligned with two gaps. The first gap is an offset of 21 aa. BLAST and identities and positive are 64% and 79%, and the first gap, an offset of Chick at the N terminal, is easily confirmed by the alignment of edges 1-5 in Fig. 6. However, near 850 BLAST finds an offset of 15 aa for Human. This leaves edges 6-8 all misaligned. Removing this offset, and replacing it with a Chick offset of 6 aa aligns all three 6-8 edges. This surprising result is discussed below. With these two gaps, values for Ψ(aa,11) are shown for edges1-8 in Table 1. The Ψ(aa,11) Human/Chick correlation is 83.1%, up from the BLAST values, and thus more informative.

The Human/Chick SMC6 profiles with W = 11 are shown in Fig. 7, aligned with a gap of 6 aa starting at 32, as identified by BLAST. Here the hydrocorrelation is only 0.768, which is lower than for other SMC Human/Chick values. The hinge region is enlarged in Fig. 8, where once again SMC6 Chick shows two hydrophobic edges spaced ~ 70 aa apart. HumSMC6 shows three closely spaced hydrophobic edges spanning 40 aa, the most refined hinge structure in the SMC family.

Overall SMC6 has evolved to be much changed from Chick to Human. This suggests that our method could yield good results for the much smaller differences between Human and Mouse. Blast identities (90.3%) and positives (96.6%) are large. The W = 11 hydrocorrelation is 96.4% The Hum/Mous variance ratio has two maxima, the first at W = 11(1.145), and the second at W = 99 (1.203), with a minimum at W = 33 (1.083). Note that the ratios of the min/(first max) and (second max)/min are both 3.0. This looks like either numerology, or a deep relationship that could be hard to prove (like Fermat's Last Theorem, conjectured in 1637, and first proved in1994).

Thus we show in Fig. 9 (W = 33) and Fig. 10 (W = 99) the Human and Mouse SMC6 profiles. For W = 33 the Human-Mouse hydrocorrelation is 0.969, and for W = 99 it is 0.990. To interpret the differences between these profiles, we can notice that the variance maxima for Human/Mouse are W = 99, larger than W = 11, larger than Mouse/Human at W = 33. Thus the in/out pattern for Mouse is better than Human at W = 33, but this is much less effective than Human at W = 11 and especially W = 99. The superiority of Human at W = 11 is evident from the fine structure of the hinge region discussed in Fig. 8, but how about W = 99 in Fig. 11?

If one looks closely at the hinge region in Fig.11, one can see that the Human peak has an unusually linear flat-top structure with steep linear sides. It is nearly linear from the minimum near 430 up to 540, then is nearly flat until 580, and then is again nearly linear down to 680. This structure creates the variance maximum for Human/Mouse at W = 99. An amusing detail here is the Human minimum near 430, sharper than the Mouse minimum. The difference is caused by a single mutation, the replacement of Mouse KKDGE at 386 by Human KKDDE at 380. Note that K, D and E are strongly hydrophilic, while Gly is hydroneutral.

On a larger scale, how has the W = 99 profile evolved from Chick to Human? This is shown in Fig. 12. Here the hydrocorrelation is 0.961. The losses of linearity in the steep hinge side 500-550, and in the hydrophilic arm 200 – 400 are large. This shows that the linearities in the SMC6 W = 99 hydroprofile are not accidental, but are instead the result of evolution improving functionality.

Discussion

Phase transition theory [6], supported by the fractal MZ hydrophobicity theory [11,8,9,2-5], has accurately predicted the evolution of Coronavirus contagiousness driven by spikes. The hydrophobicity fractals were derived from global power-law fits to water molecule accessible surface areas of > 5000 protein segment structures. The global fits extended over the range of segmental length scales $9 \leq L \leq 35$. The largest changes in CoV variance ratios occur for W ~ 35, and for most proteins the optimized W lies in the global range fitted by [11]. Because the SMC 1200 aa structure is rod-like, its sequence evolution could contain optimized values of W outside the global range, prompting the present study. DNA repair could occur over lengthy and nearly linear DNA segments, favoring linear hydroprofiles in SMC6 hinge sides and arms.

Overall the most useful value of W for SMC cohesion evolution has proved to be 57 (Figs. 1-3), a large value outside the global range used in earlier studies of many proteins. SMC condensing evolution has developed a fine structure in the hinge region, which is best resolved with W = 11, near the lower end of the global MZ range. Finally, SMC6 evolution has linearized structure on a scale of W = 99, the largest length scale encountered so far using phase transition theory. The scale used by BLAST identities and in most phylogenetic studies [15] is W = 1. The extensive sequence data in Uniprot contain a wealth of information on the evolution of protein dynamics accessible to phase transition theory [6], supported by the fractal MZ hydrophobicity results.

While protein evolutionary structural changes are usually too small to be measured structurally, even one mutation can have dramatic physiological effects, for example, in the evolution of the HPV vaccine [16]. HPV is a large capsid protein, but it was found that only the 505 aa L1 part was needed to make a good vaccine that conformationally self-assembled into morphologically correct virus-like particles (VLPs). L1 from HPV 16, taken from lesions that had not progressed to cancer, self-assembled $10^3$ times faster than the HPV 16 L1P that researchers everywhere had been using; the old strain L1P had been isolated from a cancer, which differed from L1 by only a single amino acid mutation D202H. Here the replacement of Mouse 560Pro by Human 554Leu changes the hinge fine structure of SMC6. Similarly, the deeper hydrophilic edge in human compared to mouse is caused by the replacement of KKDGE by KKDDE.

The small differences between Human and Mouse proteins have generated an industry producing humanized mice. Regeneron patented mice that have unrearranged human immunoglobulin variable regions and has used them to prepare antibody cocktails against CoV [17].

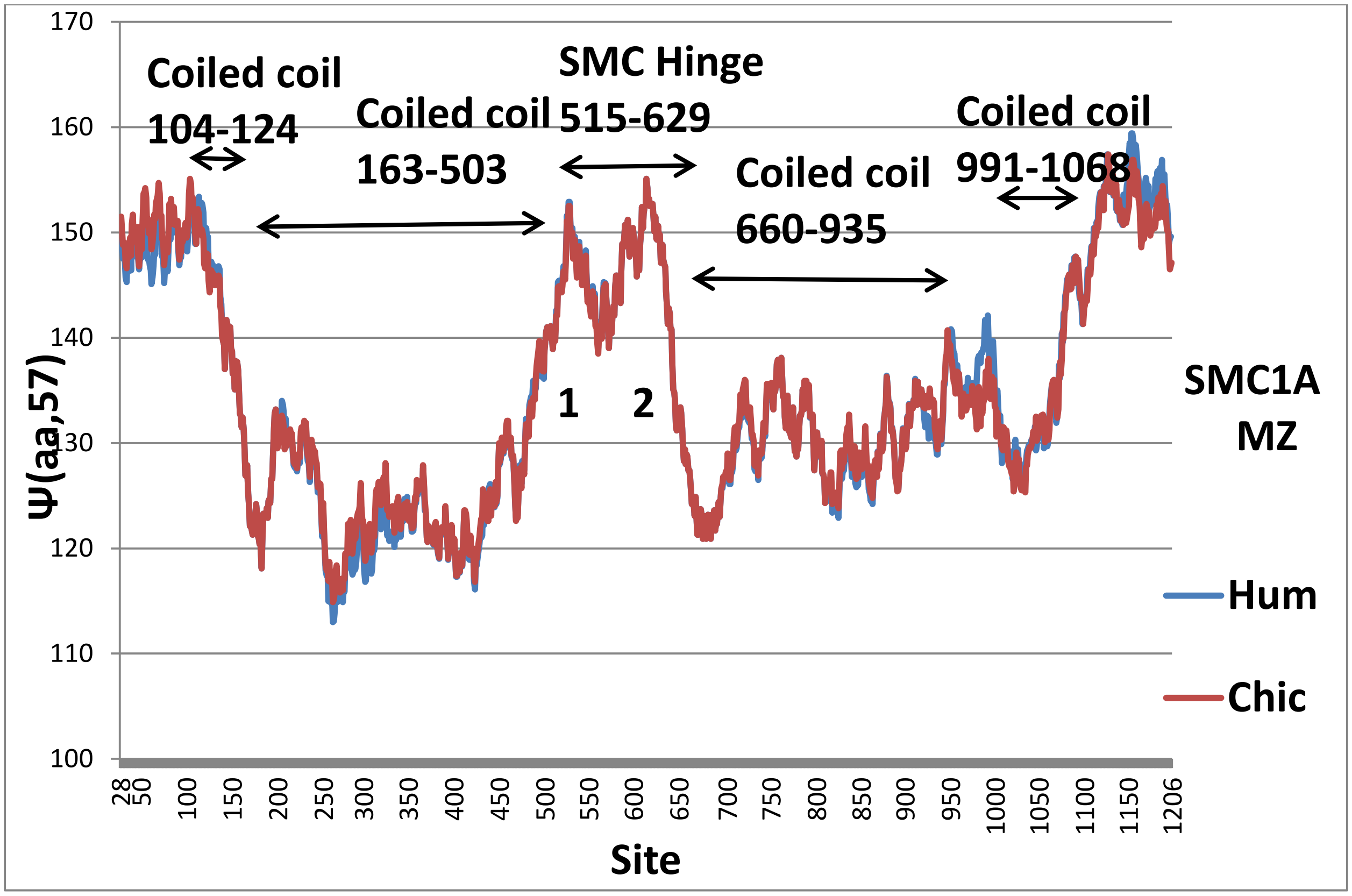

Fig. 1. The hydroprofiles of Human and Chicken SMC1A sequences, taken from and labelled as in Uniprot Q14683 and Q8AWB7. Hydroneutral is ~ 155 on the MZ scale, and the coiled – coil regions adjacent to the hinge are strongly hydrophilic, while the hinge has hydroneutral edges (1,2) separated by ~ 80 amino acids by a narrow hydrophilic region.

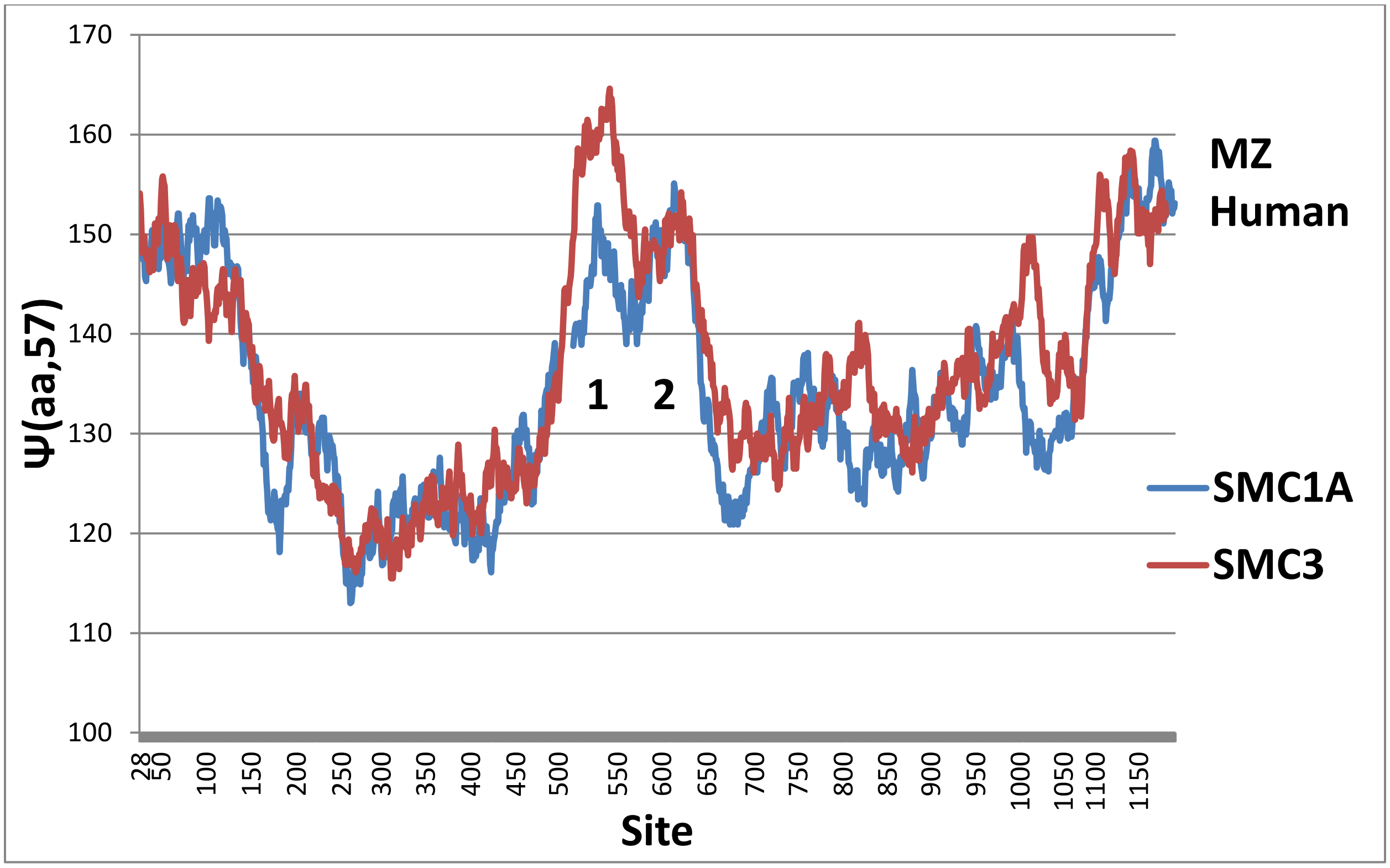

Fig. 2. There are some obvious differences between SMC1A and SMC3. The largest difference is the hinge region, where peak 1 in SMC3 is now hydrophobic. In order to align the profiles, a gap of 15 aa has been inserted into SMC1A at 503. This shows that SMC3 has buckled near 503 to produce the heterodimer rod structure (Fig. 2 of [1]).

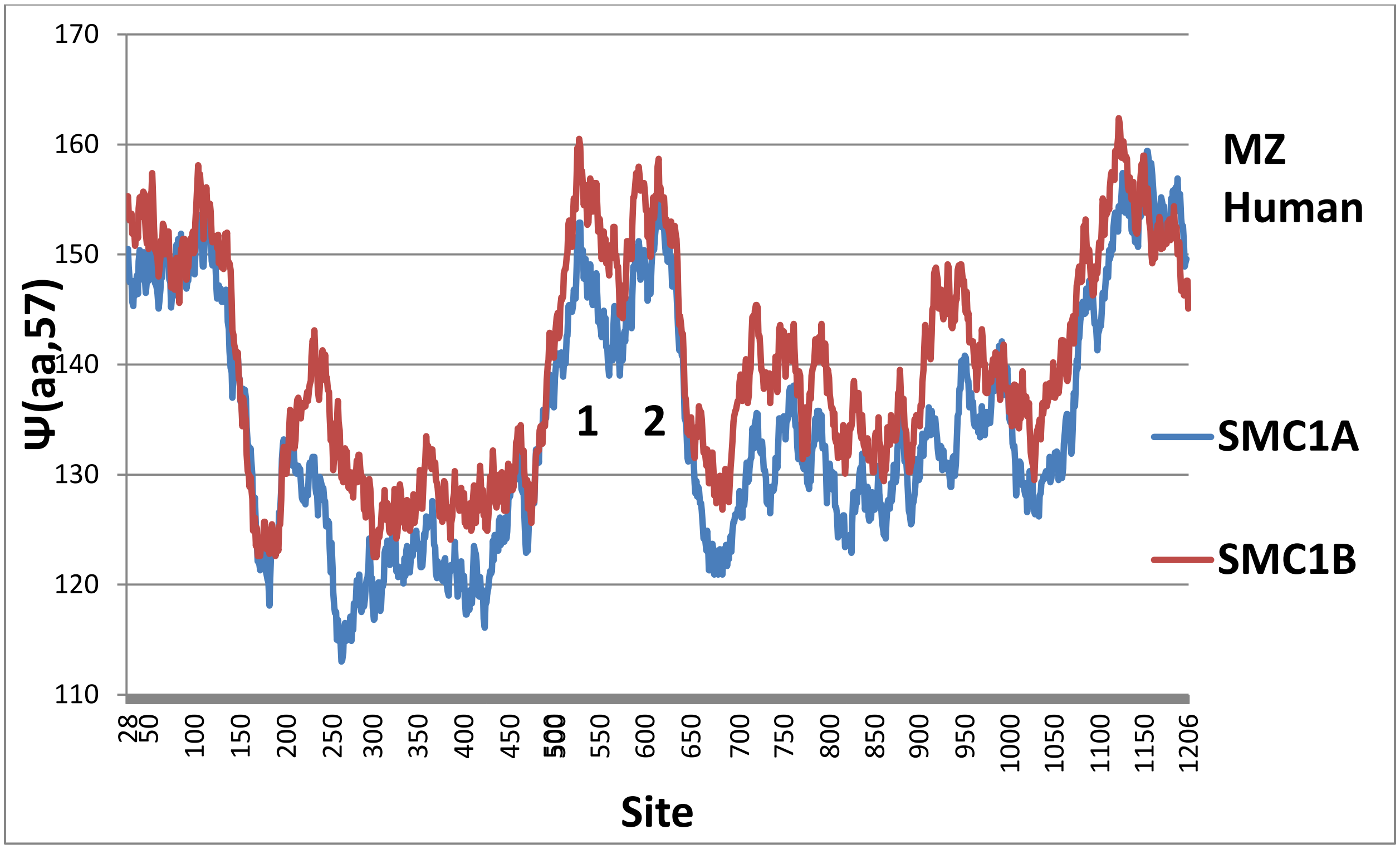

Fig. 3. Although SMC1B is quite similar to SMC1A, SMC1A is ~ 15% more hydrophilic overall. Also the relative strengths of twin hydrophobic peaks 1 and 2 are reversed here, as in Fig. 2. Only SMC1A has peak 2 more hydrophobic.

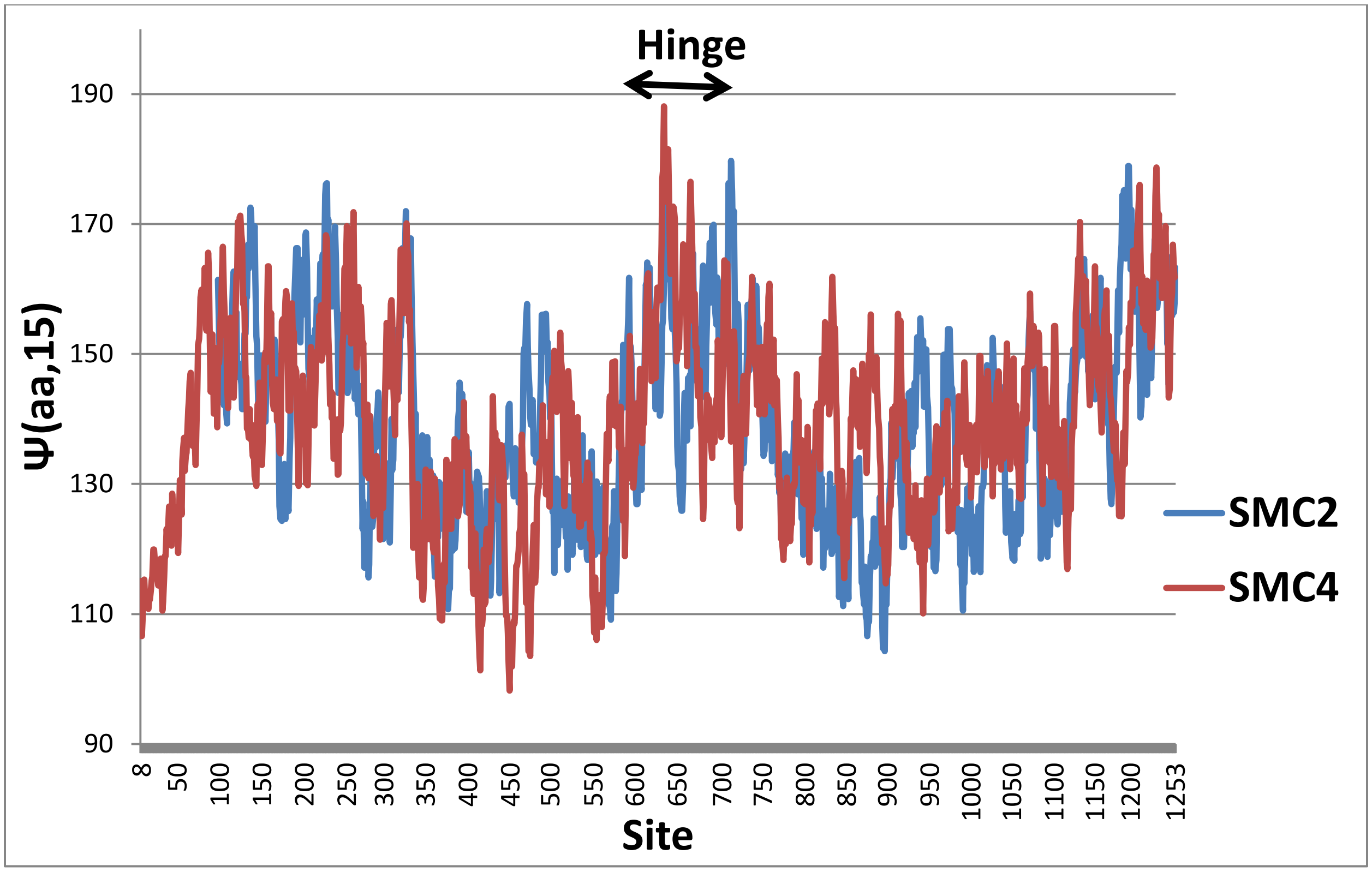

Fig. 4. For condensins compaction has reduced W from 57 in cohesions, to 15. Site numbering for SMC4,with SMC2 shifted to match hinge regions. As in Fig. 1 the coiled – coil regions adjacent to the hinge (613-727) are strongly hydrophilic. Near 450 SMC4 is more hydrophilic, while near 900 SMC2 is more hydrophilic. The complex structures of the hinge regions are enlarged in Fig. 5.

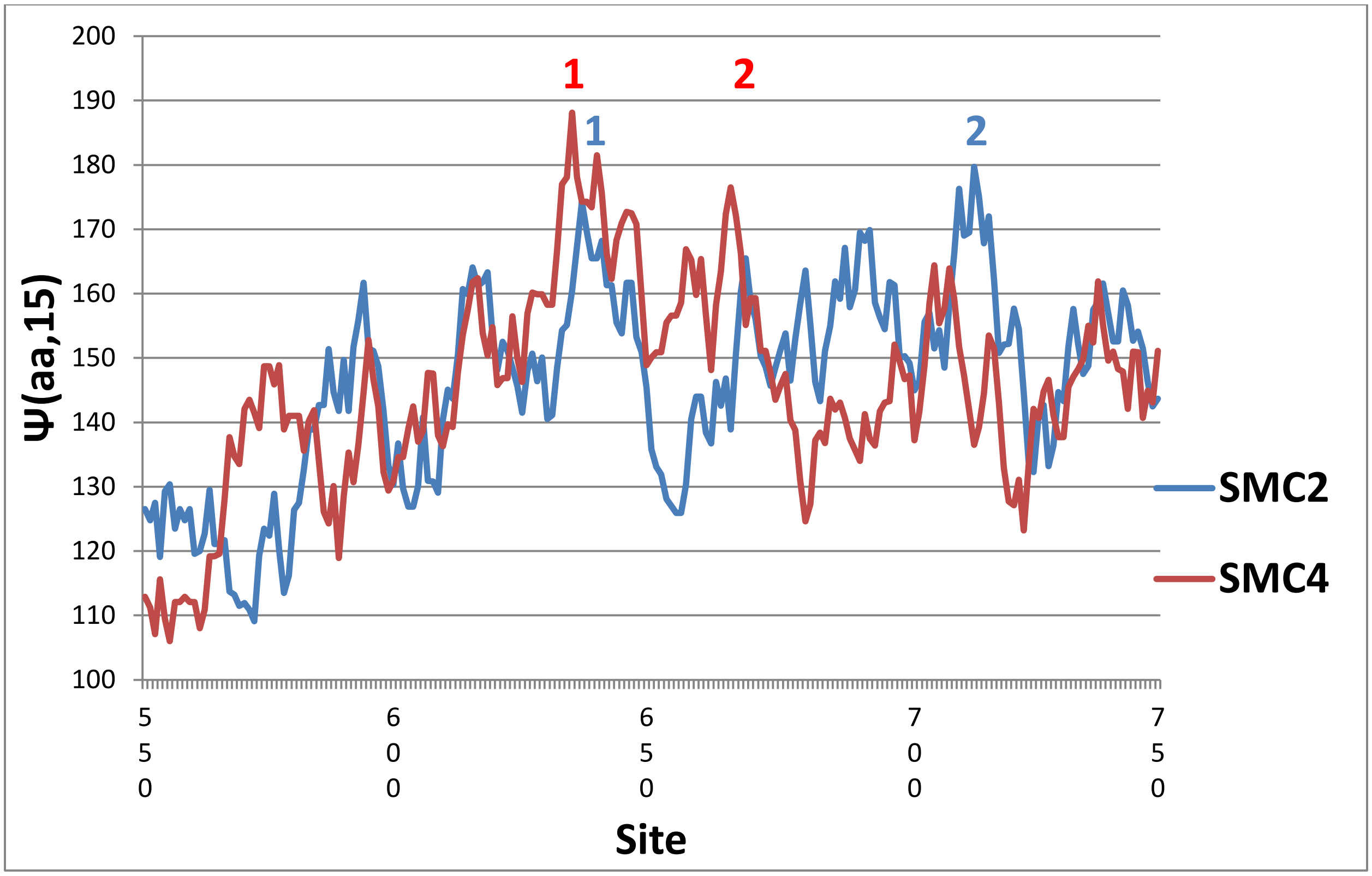

Fig. 5. Enlargement of the hinge region of condensins SMC2 and SMC4. The hinge regions of cohesions showed a common double peak structure spaced ~ 80 amino acids apart. In the condensing hinge regions (613-727) there are still two peaks, but the spacing in SMC4 has dropped to ~ 30 aa, while in SMC2 it is still ~ 80 aa.

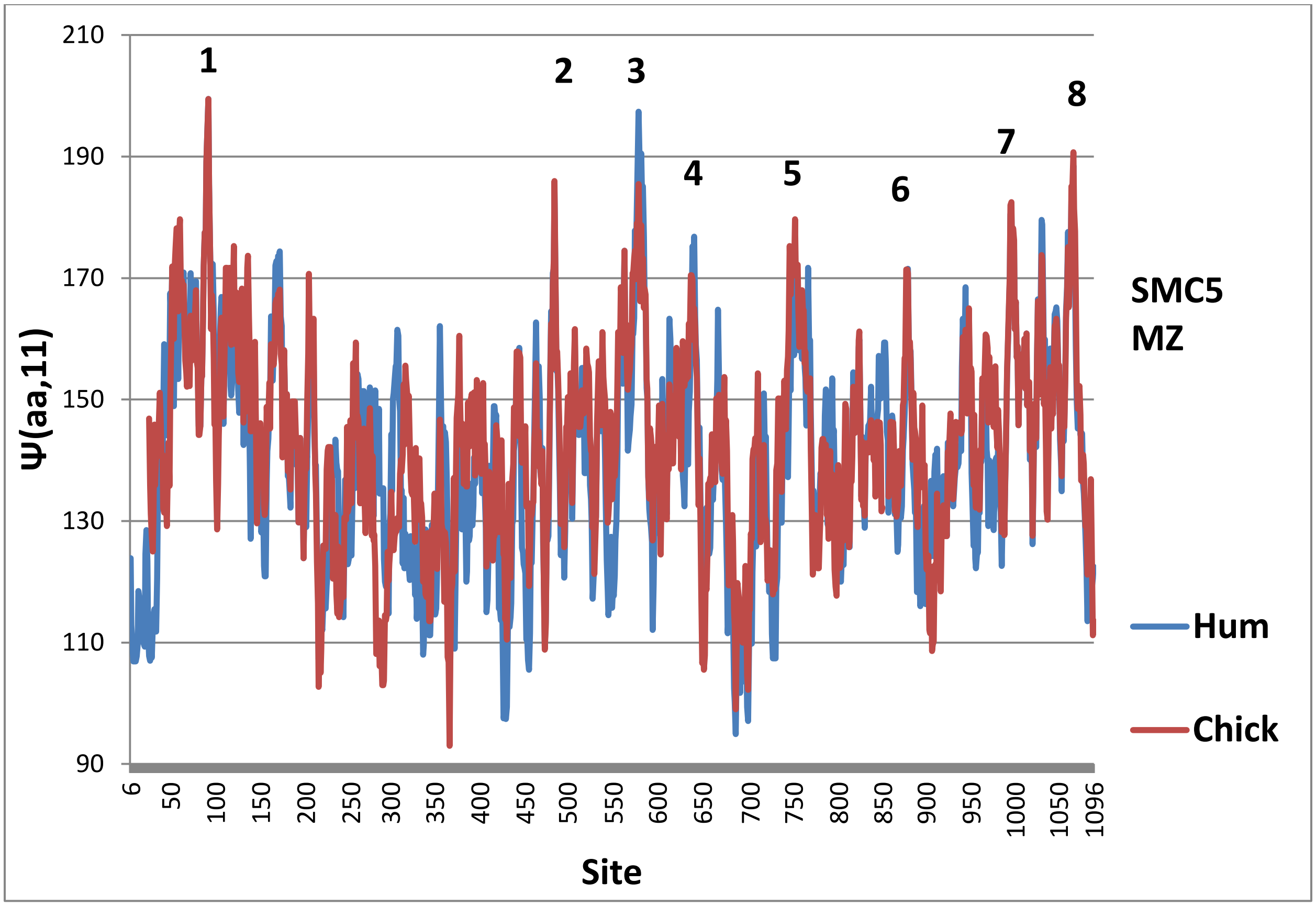

Fig. 6. Profiles for SMC5 for Human and Chick. Site numbering for Human. After offsetting for two gaps, there are at least 8 hydrophobic edges that are well conserved. The SMC5 Hinge 446-646 includes edges 2 and 3, spaced 90 aa apart, as well as edge 4. The hydrophobicities of edge 2 are nearly equal, while Human is about 10% larger for edges 3 and 4. See Table 1 for all edges.

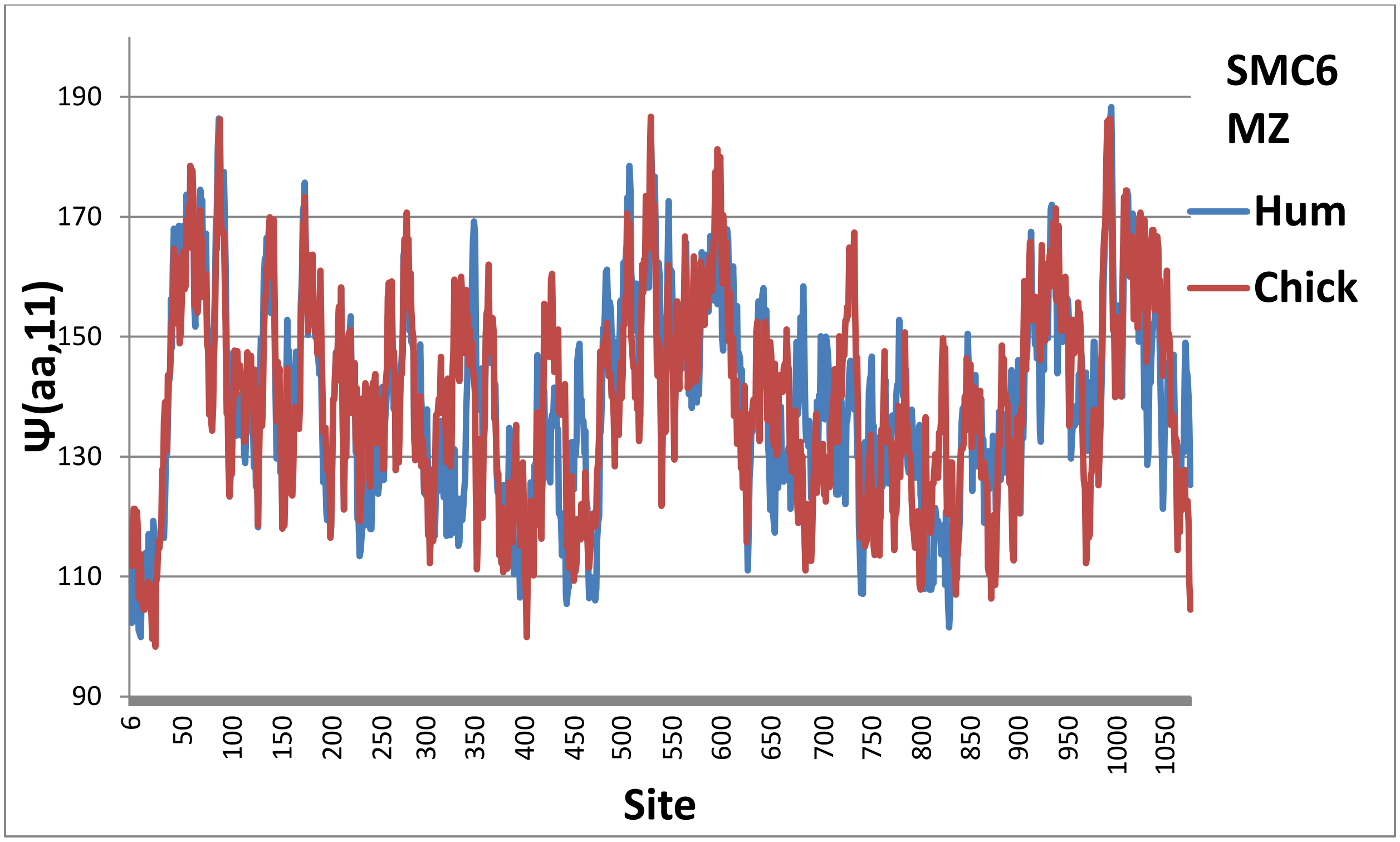

Fig. 7. Profiles for SMC6 for Human and Chick. Site numbering for Human. Here the evolutionary differences are much larger than for SMC(1-5).

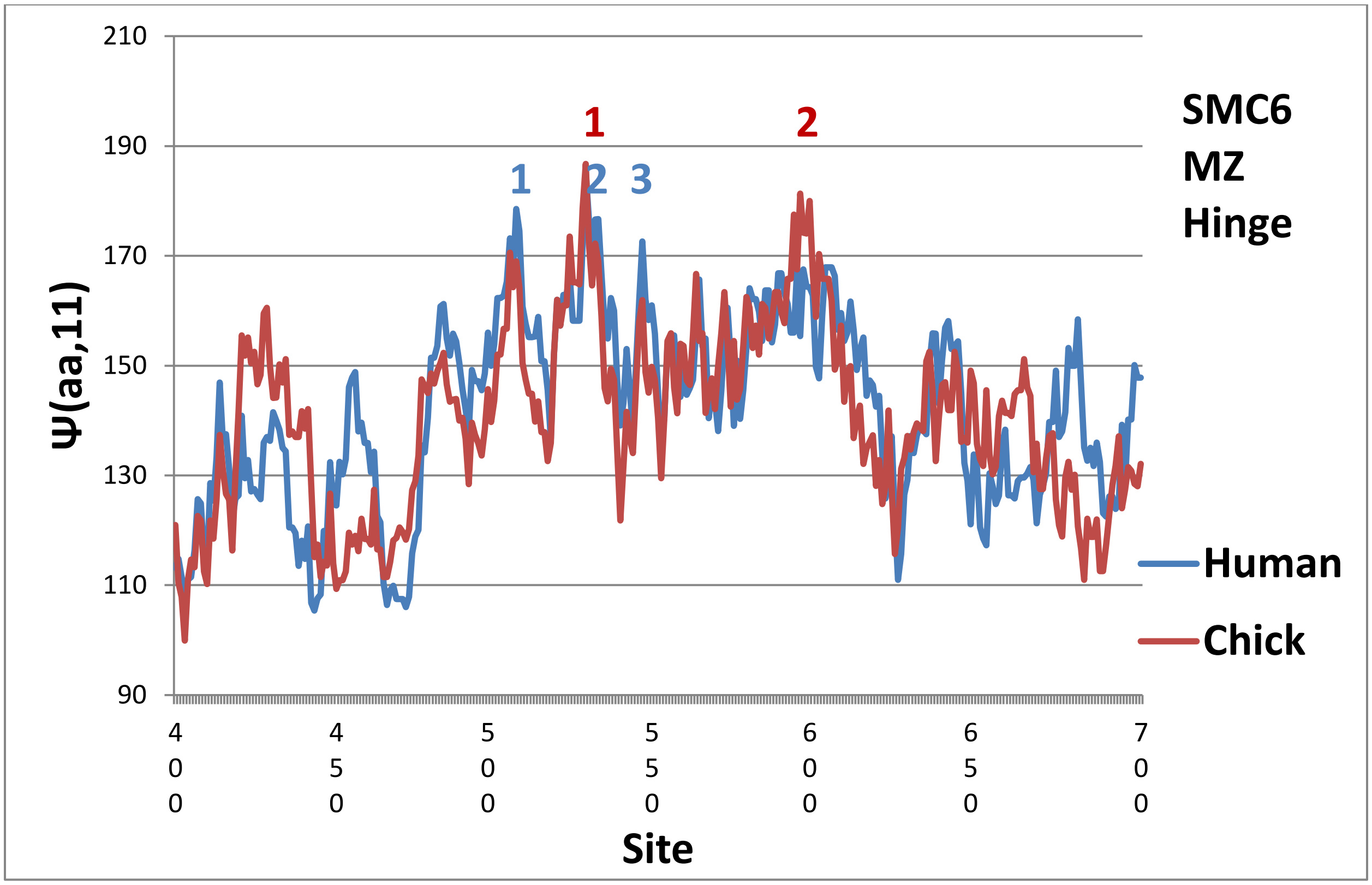

Fig. 8. According to Uniprot, the Human hinge regions extend from ~ 450 to ~ 660. The Chick hydrophobic edges (1,2) are 70 aa apart, close to the value 80 aa for SMC(1-3,5). Human SMC6 shows three closely spaced edges 1-3 spanning 40 aa.

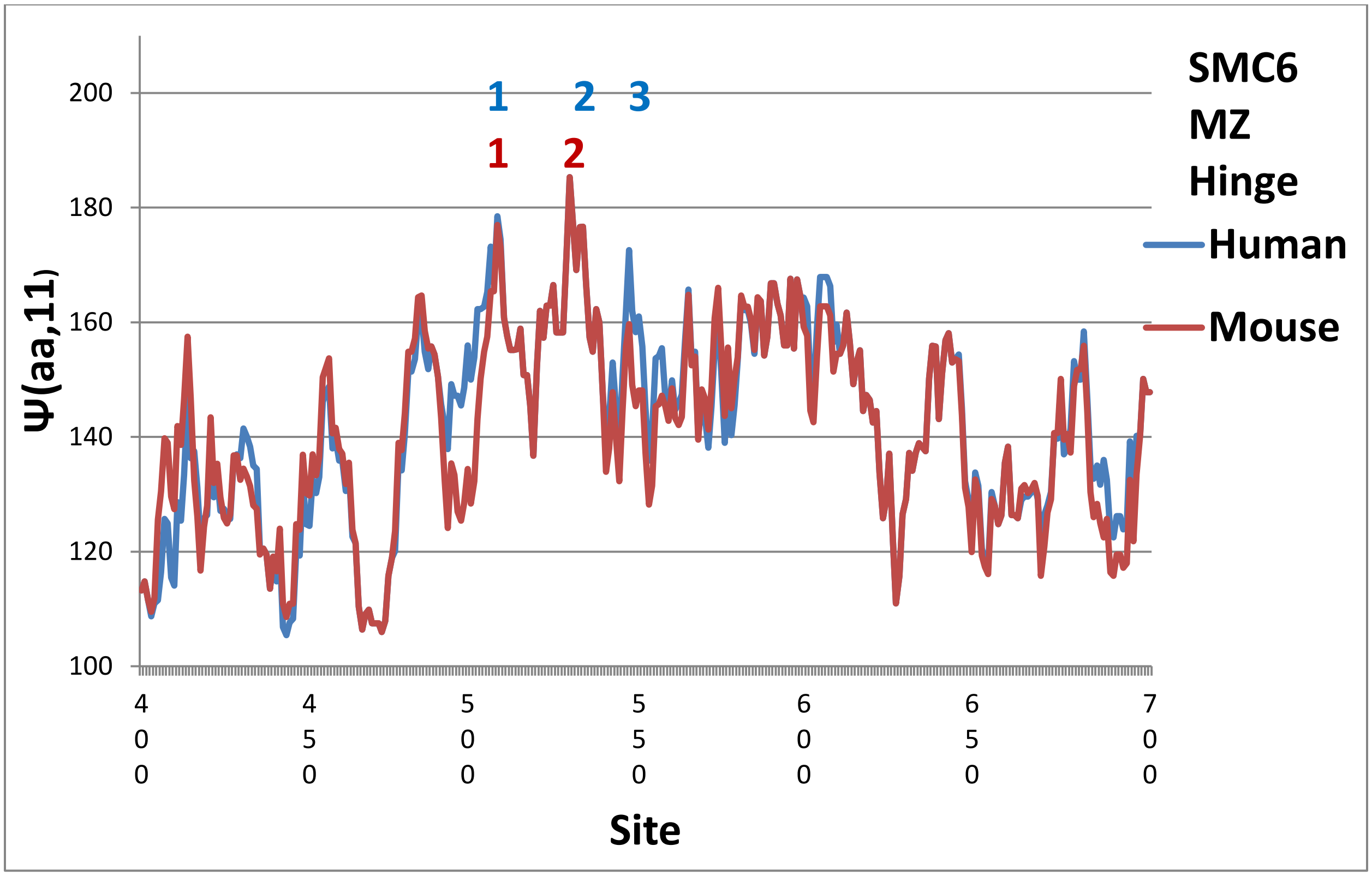

Fig. 9. The hinge regions for Human and Mouse show a surprising structure for Mouse, intermediate between Human and Chick (Fig.8). The difference in the extra Human edge 3 is caused by a single mutation, the replacement of Mouse 560Pro by Human 554Leu. There are 36 other BLAST non-positive mutations.

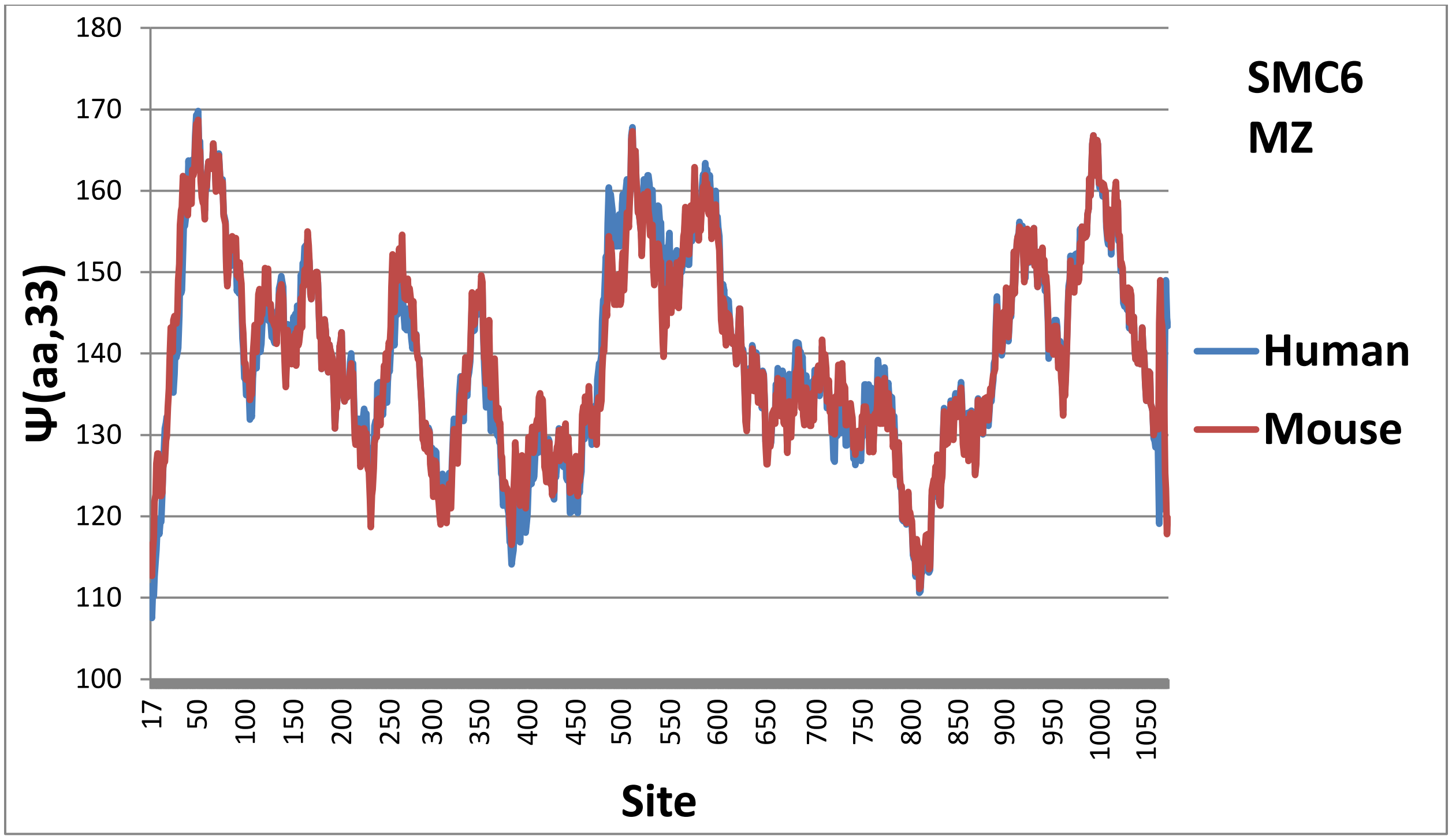

Fig.10. SMC6 profiles with W = 33. The fine structure of the Human hinge with W = 11 in Fig. 9 is not well resolved.

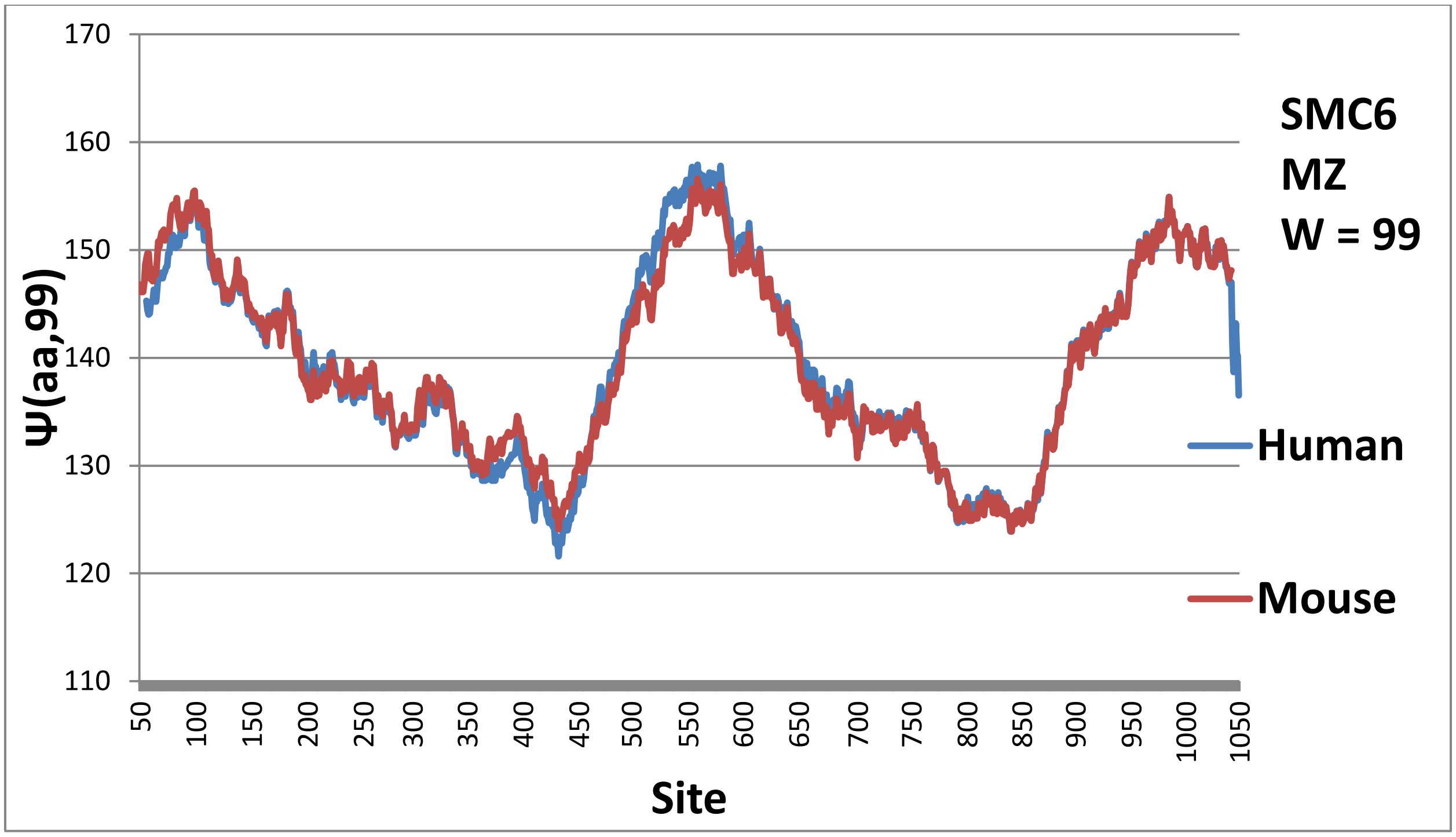

Fig. 11. The three hydrophobic edges near the N and C terminals are quite similar in Mouse and Human. The largest differences are seen in two extremities: the hinge peak near 560, which is broader and slightly higher in Human, as well as the deepest hydrophilic edge near 430, which is deeper and sharper in Human.

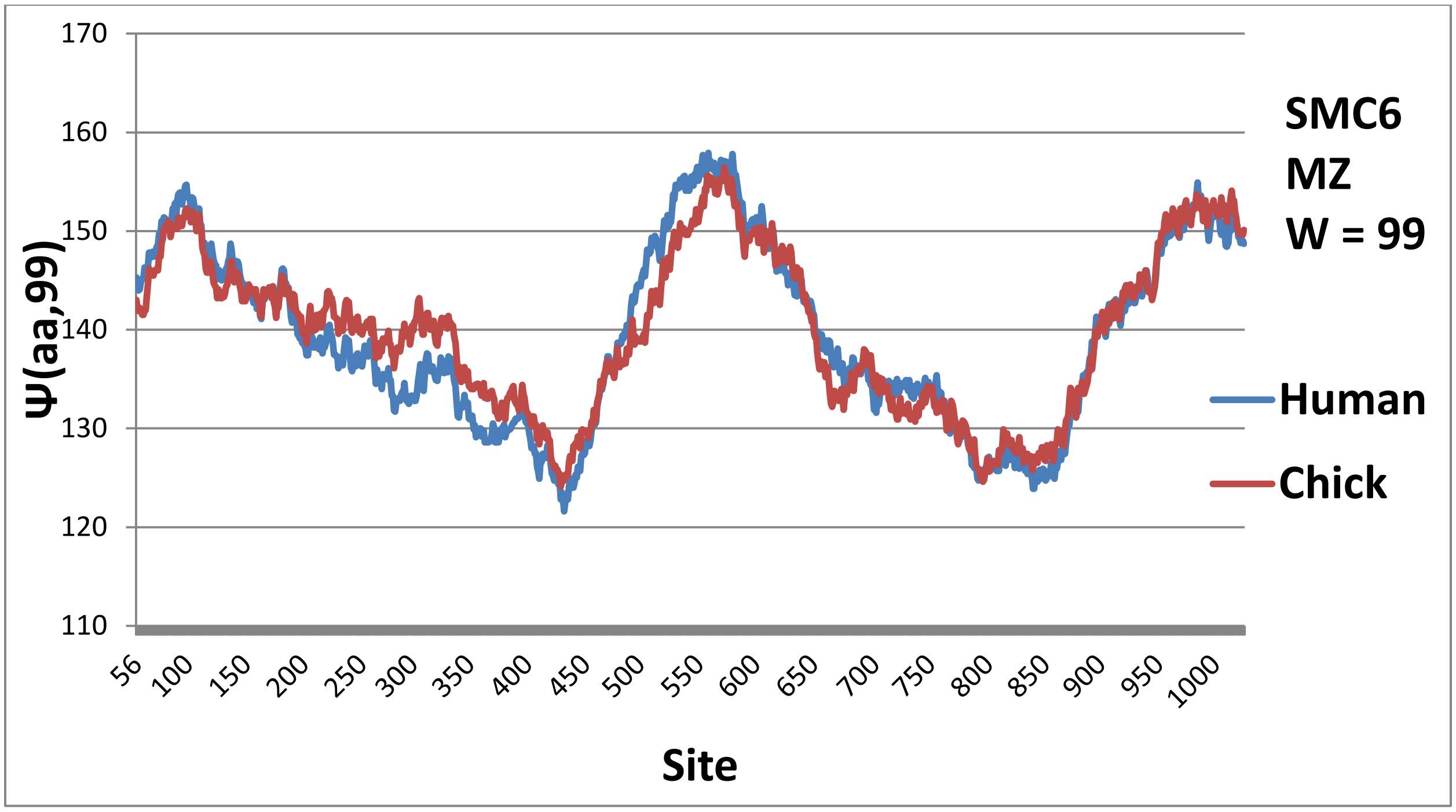

Fig. 12. Compared to Mouse in Fig. 11, here the Chick profile has lost more linearity relative to Human, both in the steep hinge side 500-550, and in the hydrophilic arm 200 – 400.

| Edge | Human | Chick |
|---|---|---|
| 1 | 199.5 | 199.5 |
| 2 | 184.5 | 186 |
| 3 | 197.4 | 185.5 |
| 4 | 176.8 | 169.8 |
| 5 | 170.5 | 179.7 |
| 6 | 171.4 | 171.5 |
| 7 | 176.6 | 178.2 |
| 8 | 188.7 | 190.5 |

Table 1. The hydrophobic edge values of Human and Chick Ψ(aa,11) are in good agreement using the two offsets discussed in the text.